\documentclass[aps,prevmat,reprint,citeautoscript,nobibnotes,superscriptaddress]{revtex4-2}

\usepackage{amsmath} 
\usepackage{graphicx} 
\usepackage{float}

\usepackage[version=4]{mhchem}

\usepackage{xcolor}
\include{tud_colors}

\usepackage{siunitx}
\usepackage{soul}

\usepackage{helvet}

\usepackage{marginnote}

\begin{document}

\title{
Understanding phase transitions of $\alpha$-quartz under dynamic compression conditions by machine-learning driven atomistic simulations
}

\author{Linus C. Erhard}
\email{erhard@mm.tu-darmstadt.de}
\affiliation{Institute of Materials Science, Technische Universit\"a{}t Darmstadt, Otto-Berndt-Strasse 3, D-64287 Darmstadt, Germany}

\author{Christoph Otzen}
\affiliation{Institute of Earth and Environmental Sciences, University of Freiburg, Hermann-Herder-Strasse 5, 79104 Freiburg, Germany}

\author{Jochen Rohrer}
\affiliation{Institute of Materials Science, Technische Universit\"a{}t Darmstadt, Otto-Berndt-Strasse 3, D-64287 Darmstadt, Germany}

\author{Clemens Prescher}
\affiliation{Institute of Earth and Environmental Sciences, University of Freiburg, Hermann-Herder-Strasse 5, 79104 Freiburg, Germany}

\author{Karsten Albe}
\email{albe@mm.tu-darmstadt.de}
\affiliation{Institute of Materials Science, Technische Universit\"a{}t Darmstadt, Otto-Berndt-Strasse 3, D-64287 Darmstadt, Germany}

\begin{abstract}
Characteristic shock effects in silica serve as a key indicator of historical impacts at geological sites. Despite this geological significance, atomistic details of structural transformations under high pressure and shock compression remain poorly understood. This ambiguity is evidenced by conflicting experimental observations of both amorphization and crystallization transitions. Utilizing a newly developed machine-learning interatomic potential, we examine the response of $\alpha$-quartz to shock compression with a peak pressure of 60 GPa over nano-second timescales. We initially observe amorphization before recrystallization into a d-NiAs-structured silica with disorder on the silicon sublattice, accompanied by the formation of domains with partial order of silicon. Investigating a variety of strain conditions enables us to identify the non-hydrostatic stress and strain states that allow the direct diffusionless formation of rosiaite-structured silica.
\end{abstract}

\maketitle

Impacts of meteorites and asteroids on the Earth have contributed significantly to its geologic evolution and the development of life. Since the formation of the proto-Earth through collisions of planetesimals in the early solar system \cite{wetherill_1986} and the subsequent formation of the Earth-Moon system through a giant impact \cite{canup_asphaug_2001}, collisions have altered the surface of the Earth through, for example, the accumulation of water \cite{albarede_2009}, sudden mass extinctions~\cite{alvarez_1980} and the creation of biological habitats \cite{osinski_2020}.

The extreme pressure and temperature conditions during such impacts lead to specific alterations in the crystal structures of rock-forming minerals known as shock effects~\cite{langenhorst_deutsch_2012}. A special role can be ascribed to the mineral quartz, which is among the most common minerals of the Earth’s continental crust and thus affected by almost every continental impact. Its lamellar amorphization and transitions to high-pressure minerals can provide reliable evidences of past impacts and estimations of peak pressures \cite{stoeffler_langenhorst_1994}. For peak pressures between 10 and 35 GPa, planar deformation features occur in shocked quartz, which represent sets of amorphous lamellae with specific crystallographic orientations. While their frequency increase and dominant orientations vary for increasing shock pressures, quartz transforms completely to a diaplectic glass for peak pressures above 35 GPa. In contrast, the thermodynamically stable high-pressure phases stishovite and coesite occur only in trace quantities \cite{stoeffler_1971}.

Transitions leading to the lamellar amorphization of quartz have previously been explained by different models. Early investigations suggested a direct transition to the amorphous state during shock compression \cite{engelhardt_bertsch_1969, grady_1980, goltrand_1992, langenhorst_1994, panero_2003}, or the transition to the thermodynamically stable phase stishovite, which would then partially transform into an amorphous solid and partially revert to quartz during decompression \cite{mcqueen_1963, ahrens_rosenberg_1968, lyzenga_1983, akins_ahrens_2002}. 
However, the lamellar amorphization of quartz was also discovered at static compression conditions 
\cite{hemley1988pressure,kingma1993microstructural}. 
While spectroscopic studies suggested direct amorphization of quartz during compression above approximately 16 GPa 
\cite{hemley1987pressure,hazen1989high,kingma1993microstructural,williams1993high,petitgirard2022anomalous}, X-ray diffraction indicated transitions to one or several crystalline high-pressure phases 
\cite{kingma1993new,kingma1996synchrotron}. A direct transition of quartz to stishovite has been excluded due to the large energy barrier associated to this generally reconstructive transition 
\cite{hainesCrystallinePostQuartzPhase2001}. Instead, various metastable phases with different crystal structures have been suggested to explain the high-pressure transitions of quartz to date~\cite{liuNewHighpressureModifications1978, sekineFe2NtypeSiO2Shocked1987, badroTheoreticalStudyFivecoordinated1997, dubrovinskyExperimentalTheoreticalIdentification1997, teterHighPressurePolymorphism1998, wentzcovitchNewPhasePressure1998, hainesCrystallinePostQuartzPhase2001, choudhuryInitioStudiesPhonon2006, martonakCrystalStructureTransformations2006}.

\begin{figure*}[t]
    \centering
    \includegraphics{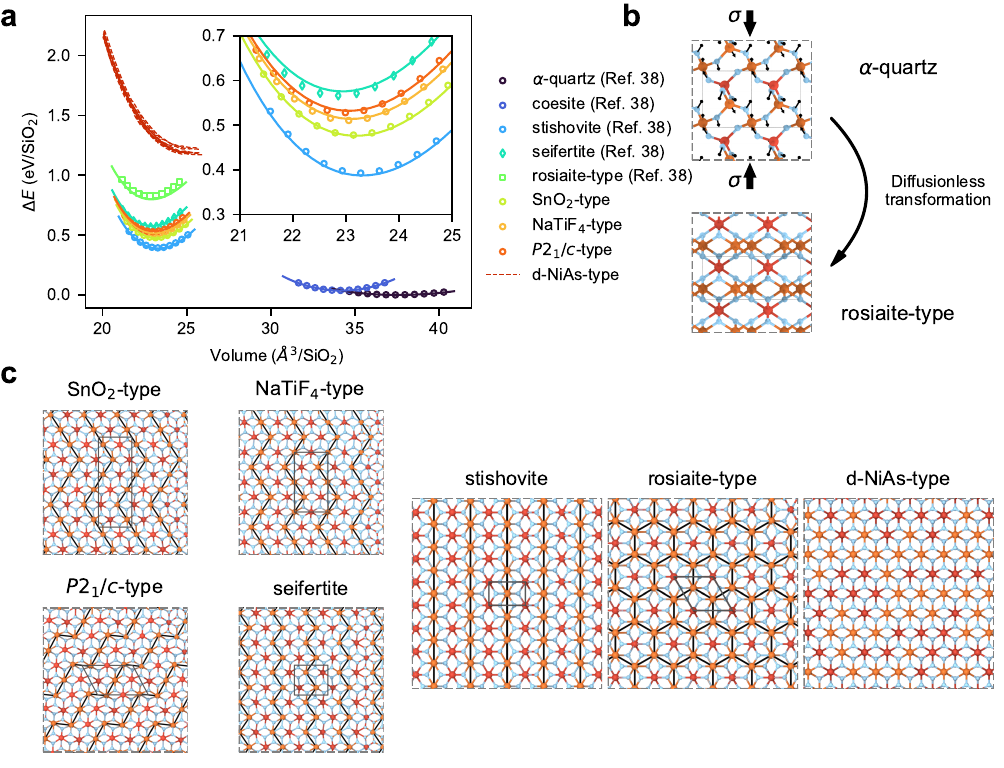}
    \caption{(a) Energy-volume curves of various high-pressure polymorphs of silica compared to the energy-volume curves of $\alpha$-quartz. The lines correspond to the prediction by the ACE potential, while the points correspond to DFT (SCAN) results. In case of $\alpha$-quartz, coesite, stishovite, seifertite and rosiaite the data is taken from Ref. \citenum{erhard2024modelling}. Since in the case of d-NiAs the silicon atoms are randomly distributed, we show the energy-volume curves of 10 different structures with each 3000 atoms. Due to the system size, we show in this case only the ACE result. (b) Illustration of the direct transformation of $\alpha$-quartz to rosiaite-type silica, which was already suggested by Tsuchiya and Nakagawa \cite{tsuchiyaNewHighpressureStructure2022}. The black arrows indicate the necessary displacements of the atoms. Blue atoms indicate oxygen atoms, while orange and red atoms indicate silicon atoms of different layers in the rosiaite-type structure. (c) Various silica polymorphs, which are all based on a hcp oxygen sublattice. The differences between the structures result from the distribution of silicon atoms on these structures. In case of stishovite, the silicon atoms are arranged on straight lines. In contrast to this for SnO$_2$-, NaTiF$_4$-, $P2_1/c$-type silica and seifertite silicon atoms are arranged on different zig-zag patterns. Rosiaite-type silica has instead one silicon layer, which has a honeycomb-like structure similar to graphene. In case of the d-NiAs structure, the silicon atoms are not ordered, but instead are randomly placed in octahedral voids with a probability of 50\%. Color coding of the atoms is identical to (b).}
    \label{fig:overview}
\end{figure*}

Among the numerous high-pressure silica polymorphs, the most stable structures can be described by a hexagonally closest packed (hcp) arrangement of oxygen, in which silicon occupies half of the available octahedral interstices \cite{teterHighPressurePolymorphism1998}. The numerous possibilities to distribute silicon in these interstices gives rise to a large number of polymorphs that are characterized by different silicon orderings. The crystal structures and energy volume curves of such high-pressure polymorphs are shown in Fig. \ref{fig:overview}.

X-ray diffraction patterns obtained during shock compression of $\alpha$-quartz indicate the formation of the defective niccolite (d-NiAs) type structure \cite{tracyStructuralResponseAquartz2020}, which was also reported in shock recovery \cite{sekineFe2NtypeSiO2Shocked1987} and heated diamond anvil cell \cite{liuNewHighpressureModifications1978, dubrovinskyClassNewHighpressure2004a, prakapenka2004high} experiments.
This structure is characterized by a random distribution of silicon in octahedral interstices of the hcp arrangement of oxygen, and might be formed due to the short time scales of shock compression suppressing silicon diffusion into ordered positions~\cite{tracyStructuralResponseAquartz2020}.
An alternative interpretation of the X-ray diffraction patterns is provided by a mixture of various silica phases with different silicon ordering.

A major issue with the interpretation of laboratory shock experiments and their comparison to natural impacts is the difference of time scales \cite{decarli_2002}.
The peak pressure and load duration during natural impacts is on the order of milliseconds for small craters (Meteor Crater) up to seconds for large impacts (Chicxulub).
However, laboratory experiments are rather in the nanosecond to microsecond range, depending on the chosen technique.
Thus, the material response to shock load is highly affected by kinetic parameters and therefore might not be correctly mimicked in a laboratory setting.
A recent approach tries to assess the kinetic effects by compressing quartz rapidly using a membrane diamond anvil cell resulting in a transformation to a high-pressure polymorph with  the crystal structure of the mineral rosiaite \cite{otzenEvidenceRosiaitestructuredHighpressure2023}.
This transition was additionally reported by computational studies, revelaing a diffusionless mechanism \cite{tsuchiyaNewHighpressureStructure2022}.
The rosiaite phase provides an explanation for the amorphization of quartz due to its instability at ambient conditions causing a collapse to an amorphous solid during decompression \cite{otzenEvidenceRosiaitestructuredHighpressure2023}. Moreover, the formation of rosiaite-structured silica during compression of quartz can also serve as an intermediate step for subsequent solid state nucleation of stishovite inside the rosiaite-structured phase \cite{otzenNewMechanismStishovite2024}.
The very similar microstructures between the quartz samples recovered from these experiments and naturally shocked quartz samples suggests, that the rosiaite-structured phase might also be responsible for the lamellar amorphization and solid-state stishovite formation during impacts observed in natural samples.

\begin{figure*}[p]
    \centering
    \includegraphics[width=0.9\textwidth]{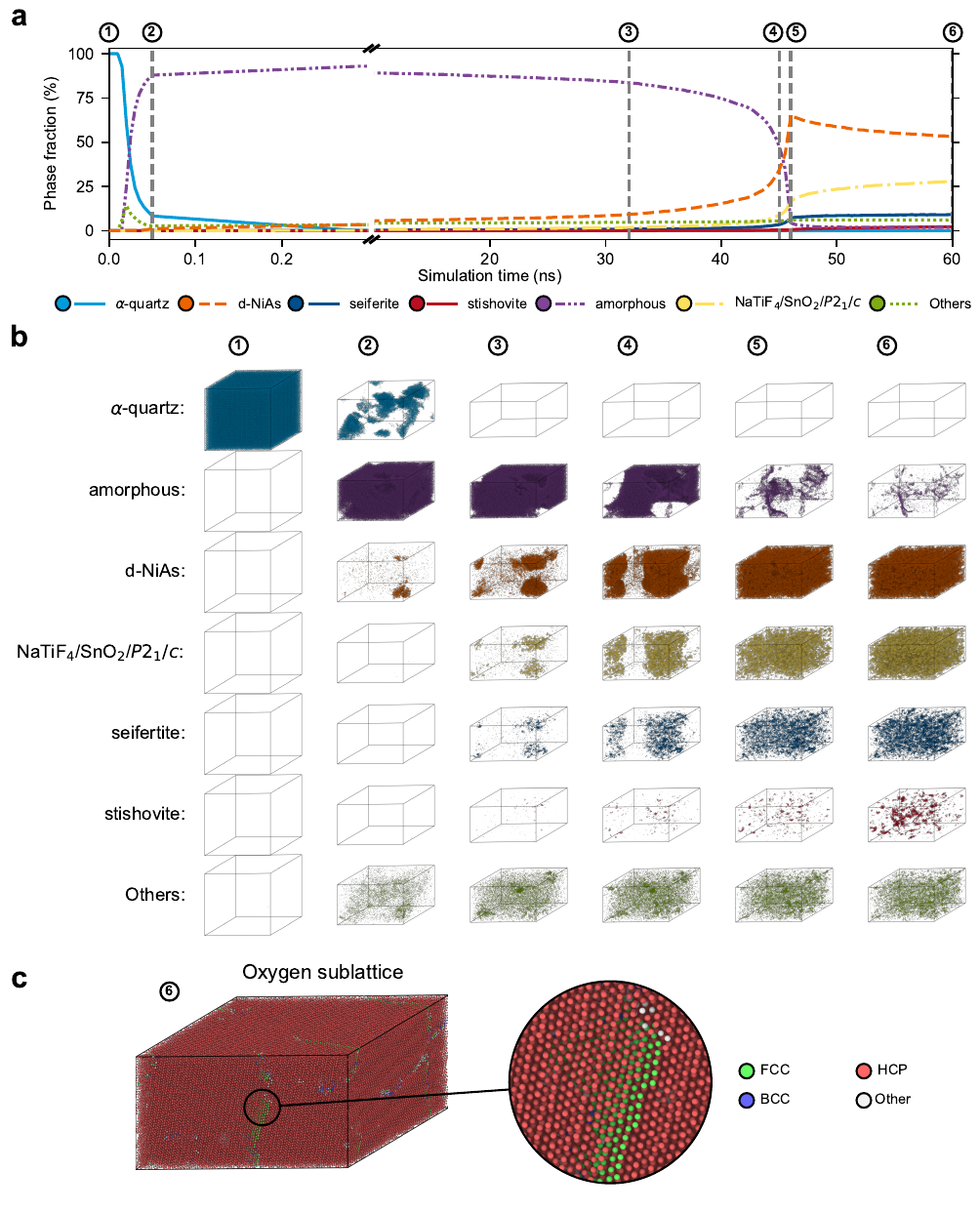}
    \caption{(a) Dynamic graph-convolutional neural network (DG-CNN) structure identification results for a shock simulation of $\alpha$-quartz under a pressure of 56~GPa. The phase fraction implies how many atoms are identified with a certain structure at a certain time step. All atoms which are classified as NaTiF$_4$/SnO$_2$ or $P2_1/c$-type silica are summarized in one category. Atoms, which have been identified to be any other not mentioned crystal structure, are summarized as \textit{Others}. (b) Snapshots from the shock simulation showing only the atoms with the corresponding DG-CNN identified structure in each row. Atoms are color coded according to the structural classification from the DG-CNN structure identification algorithm. Colors are the same as in (a). Temperatures during the simulation are shown in Supplementary Fig. S2. (c) Polyhedral template matching analysis (RMSD: 0.2) result for the oxygen sublattice of the structural snapshot after 60~ns simulation time. Nearly the whole structure is classified as hcp. One sees that quartz is first amorphizing and then recrystallizing within the d-NiAs structure with locally deviating  arrangements of silicon atoms.}
    \label{fig:shock}
\end{figure*}

In the present study, we perform computer simulations in order to investigate the transitions of quartz observed in the two above described experiments. 
To understand the experiments conducted in \cite{tracyStructuralResponseAquartz2020} molecular dynamics shock simulations of $\alpha$-quartz using the Hugoniostat method \cite{ravelo2004constant} are carried out. Further, we vary the strain state of $\alpha$-quartz systematically and analyze possible strain-dependent phase transitions using a combination of molecular statics and molecular dynamics simulations to reconcile the results of \cite{otzenEvidenceRosiaitestructuredHighpressure2023}.
Lastly, we perform solid-state nudge elastic band (NEB) calculations to analyze the transition paths of $\alpha$-quartz. In contrast to earlier studies, our simulations are based on a machine-learning interatomic potential, which was fitted to high quality density-functional theory (DFT) data. This potential shows accurate transition pressures for the high-pressure silica polymorphs, is able to accurately describe the high-pressure amorphous phase, and has been shown to be accurate for silica polymorphs that were not part of the training dataset \cite{erhard2024modelling}. Therefore, the machine-learning interatomic potential allows a more accurate description of the high-pressure phase transitions of quartz as compared to previously employed classical potentials. Moreover, we address the complex problem of structure identification through a novel structure identification model based on dynamic graph convolutional neural networks (DG-CNN) \cite{erhardutt2024structure}.

\section*{Results and Discussion}

\subsection*{Shock simulation of $\alpha$-quartz}

As a first step we performed an uniaxial molecular dynamics shock simulation along the c-axis of $\alpha$-quartz replicating the shock experiments by Tracy \textit{et al.}~\cite{tracyStructuralResponseAquartz2020} with a peak pressure of 56 GPa. 
While the characteristic time scales often induce strong challenges for investigating the material response with existing methods experimentally, these time scales are on the same order of magnitude as we can simulate in a MD simulation. The summary of the simulation results is shown in Figure \ref{fig:shock}.

At the beginning nearly all atoms are classified as $\alpha$-quartz atoms. 
However, the fraction of $\alpha$-quartz classified atoms rapidly decrease already in the first 50~ps with only few domains left (See also Fig. \ref{fig:shock}b).
After 300~ps all traces of $\alpha$-quartz disappear. 
Simultaneously, the phase fraction of the amorphous phase is strongly increasing.
At 300~ps the amorphous phase fraction reaches a value of roughly 90\%. 
Next to the amorphous phase already in the initial phase of the shock simulation another phase namely the d-NiAs type SiO$_2$ appears. 
This phase consists of a hcp oxygen sublattice and randomly distributed silicon atoms. 
Over the next 30~ns the phase fraction of the d-NiAs type silica phase is increasing, while the fraction of the amorphous phase is decreasing. 
This can be also seen in the snapshots in Fig. \ref{fig:shock}b between 50~ps and 41~ns. 
Between 40 and 50~ns the growth of the d-NiAs-phase is strongly increasing leading to a nearly completely crystallized sample at 60~ns (see last snapshot Fig. \ref{fig:shock}b).
Polyhedral template matching results shown in Fig. \ref{fig:shock}c demonstrate that the oxygen sublattice of this sample is a hcp lattice with only few defects.
However, due to the different silicon ordering we can still have different types of crystal structures (see Fig. \ref{fig:overview}b). 
Indeed, after a peak between 40~ns and 50~ns the phase fraction of the d-NiAs phase is decreasing again. 
This goes along with an increase in the fraction of other crystalline phases, while the fraction of the amorphous phase stays negligible.
Especially, more and more small domains appear, which are classified to be NaTiF$_4$-type, SnO$_2$-type and $P2_1/c$-structured silica as well as seifertite.
Nevertheless in more than 50\% of the matrix the d-NiAs phase is still the dominant phase. 
Moreover, stishovite, which should be the stable phase at these conditions, happens to have a negligible phase fraction, which growths much slower than e.g. the phase fraction of seifertite (see also Fig. \ref{fig:shock}b). 
Also rosiaite-structured silica, which has been observed in static compression experiments \cite{otzenEvidenceRosiaitestructuredHighpressure2023,otzenNewMechanismStishovite2024}, does not appear in significant amounts.

\begin{figure}[t]
    \centering
    \includegraphics{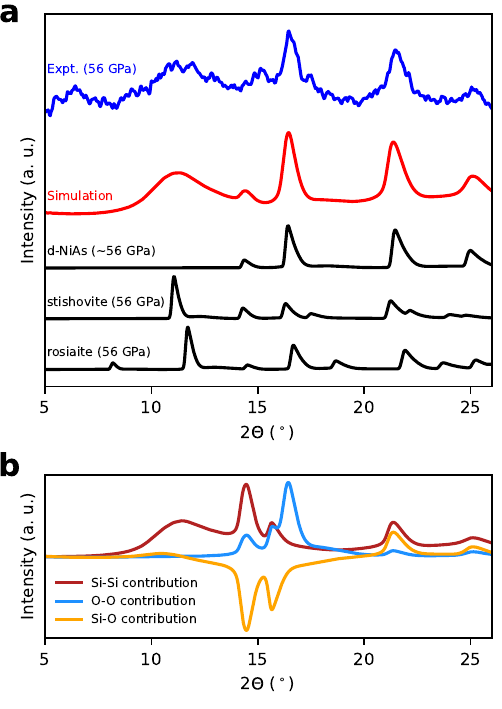}
    \caption{(a) Calculated XRD pattern of a shock simulation of $\alpha$-quartz after 60~ns compared to the experimental XRD pattern measured \textit{in situ} by Tracy \textit{et al.} \cite{tracyStructuralResponseAquartz2020}. Both experiment and simulation have been performed under 56~GPa with compression along the c-axis of $\alpha$-quartz. Additionally, the theoretical pattern for stishovite, rosiaite and d-NiAs are shown in comparison to the simulation result and the experimental data, with cell adjusted to the same pressure. In case of d-NiAs we used the average lattice parameters of the structures from Fig. \ref{fig:overview}a at 56~GPa. For all calculations we used the same X-ray wavelength distribution as it was reported in the work by Tracy \textit{et al.} \cite{tracyStructuralResponseAquartz2020}. (b) Individual contributions of the Si-Si distances, the O-O distances and the Si-O distances to the total XRD pattern of the shocked structure after 60~ns.}
    \label{fig:xrd}
\end{figure}

In experimental X-ray diffraction (XRD) measurements of shock impacts on $\alpha$-quartz the resulting measured structure was determined to be d-NiAs-type silica~\cite{tracyStructuralResponseAquartz2020}. 
This is in good agreement to our structure identification results. 
Moreover, the timescales of the measurement are with 100~ns similarly to the timescales of our simulation. 
To allow for better comparison to experiment, we calculate the XRD pattern of our simulation cell after 60~ns simulation time and compare it to the experimental diffraction pattern in Fig.~\ref{fig:xrd}a. 
Indeed, we are able to observe very good agreement between experimental XRD pattern and the XRD pattern of the simulation.
Also in both cases the XRD pattern agree well with the theoretical XRD pattern of the d-NiAs phase, while characteristic peaks of stishovite, rosiaite and several other phases are missing (see Supplementary Fig.~S1). 
Instead of these characteristic peaks at low angles, there is a diffuse peak between a 2$\Theta$ of 10$^\circ$ to 13$^\circ$ similar to the first sharp diffraction peak peak of amorphous silica. 
Therefore, one assumption of the origin of this peak has been that the sample is partially amorphised \cite{tracyStructuralResponseAquartz2020}. We can clearly exclude this, since the resulting structure from our simulation is crystallized to nearly 100\%. 
Instead, we assume that this broad feature is caused by localized partial ordering of silicon atoms.
In case of the perfect d-NiAs structure the silicon is distributed completely random on the octahedral positions. 
Therefore, lacking the low angle peak caused by ordered silicon. 
Energetically, however, a totally random distribution of silicon would be highly unfavourable. 
Moreover, in our simulation significant parts of the structure are identified as seifertite, NaTiF$_4$-type, SnO$_2$-type and $P2_1/c$-type silica.  
All these structures exhibit sharp peaks in the diffraction angle range of the broad feature. 
Since clusters of these structures are spatially limited no long range order exist.
Instead, the silicon exhibts a certain short and intermediate range order. 
This is also supported by the contributions of different interactions to the simulated XRD pattern, which are shown in Fig.~\ref{fig:xrd}b. 
Indeed it can be seen that the broad peak is mainly influence by the silicon-silicon interactions. 
We note here, that our current shock simulations are in contradictions with earlier shock simulation of quartz, which found stishovite after shock~\cite{shenNanosecondHomogeneousNucleation2016}.
The excellent agreement with experiment, here, shows the high quality of the used machine-learning interatomic potential, giving more accurate insights compared to earlier classical interatomic potentials.

\begin{figure*}[t!]
    \centering
    \includegraphics{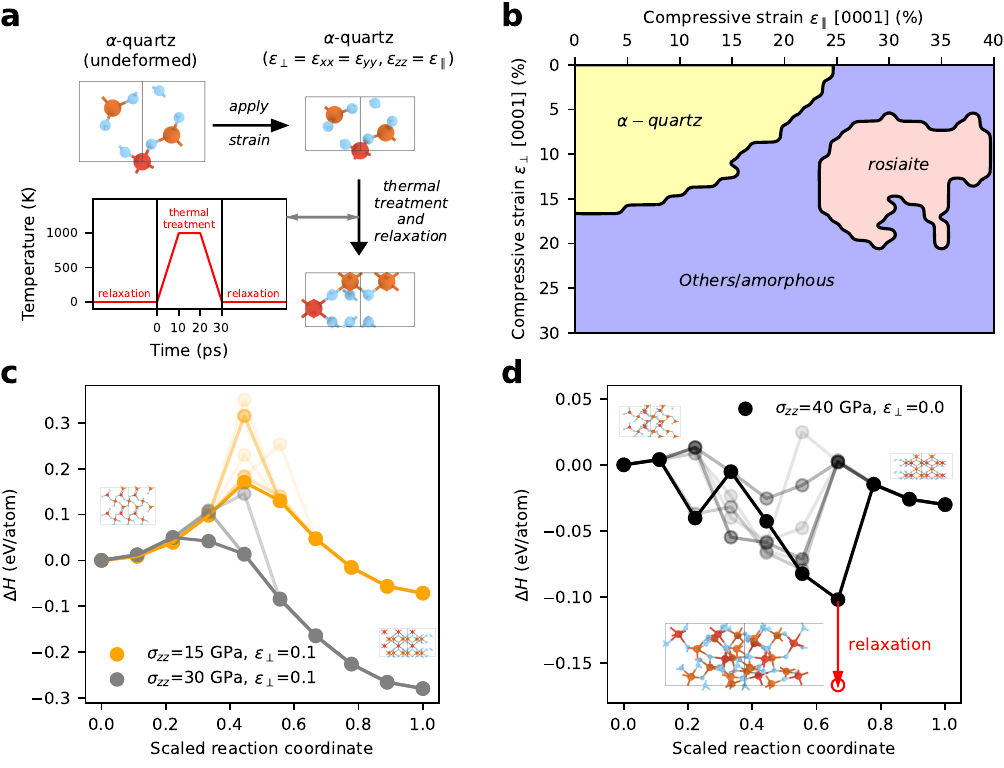}
    \caption{(a) An undeformed $\alpha$-quartz cell has been used as input for a large number of deformations. These include strains between 0 to 30\ \% in the $a$-$b$ plane of $\alpha$-quartz and 0 to 40 \% along the $c$-direction of $\alpha$-quartz. After applying the strain to the cell, the structure is undergoing a thermal treatment for 30~ps with temperatures up to 1000~K and several energetical minimizations. This was done for supercells up to 8x8x8 size. (b) The phase-stability diagram shows the result of various compression simulations of $\alpha$-quartz and the arising structure after the minimization and thermal treatment. Basis of the plot are dynamic graph convolutional neural network structure identification results. Due to the dominance of rosiaite in the structure identification results, we focus on this structure, although also other structures have been detected in smaller amounts (see Supplementary Fig. S3, S4 and S5. (c) Transition path between $\alpha$-quartz and rosiaite calculated by solid-state nudge elastic band method. We set the lattice parameter of $\alpha$-quartz as well as rosiaite in $a$-$b$-direction to the same value. This is 90\% of the undeformed $\alpha$-quartz value, or 10\% compressively strained $\alpha$-quartz. The $c$ lattice parameter is determined by minimization along the $c$ direction under an applied pressure of 15~GPa or 30~GPa. This results in a compressive strain of 4.1\% (15~GPa) and 8.2\% (30~GPa) for quartz and 27.5\% (15~GPa) and 29.0\% (30~GPa) for rosiaite-structured silica in $c$-direction, in both cases referenced to the undeformed $\alpha$-quartz cell. We used different possible displacive paths, by employing several rosiaite structures from (b). The minimum energy path is shown in bold, while the other paths are shown blurred.  (d) Transition paths like in (c), however, this time the $a$-$b$ lattice parameters of $\alpha$-quartz and rosiaite-structured silica are scaled to the relaxed $\alpha$-quartz reference. In $c$-direction a pressure of 40~GPa is applied, which results in a compressive strain of 23.7\% for quartz and 36.6\% for rosiaite-structured silica in $c$-direction. Since on the transition path lower energy structures appeared, we picked one of these and relaxed it including box and positions. }
    \label{fig:phases_strain}
\end{figure*}

But why does $\alpha$-quartz transform under shock into the d-NiAs structure and not into stishovite, which is the thermodynamically stable phase under these conditions?
In the case of vitreous silica \textit{in situ} shock experiments and simulations have shown, that stishovite can crystallize under shock~\cite{gleasonUltrafastVisualizationCrystallization2015,tracySituXRayDiffraction2018a,shenNanosecondHomogeneousNucleation2016}. 
However, recent MD shock simulations of amorphous silica have found, that before silica crystallizes in the stishovite structure, an intermediated state namely the d-NiAs-structured silica can be observed~\cite{erhardutt2024structure}. 
This shows that when silica crystallizes from the amorphous phase, first the oxygen sublattice is rearranged to a hcp lattice. 
Only secondly the silicon arranges within the structure to stishovite. 
In the case of $\alpha$-quartz only the first step of this crystallization process is seen within the simulated time scales. 
One reason for this are also the significantly lower temperature during shock of $\alpha$-quartz compared to the shock of amorphous silica (see Supplementary Figure~S2). 
These higher temperatures allow the much faster arrangement of oxygen and silicon atoms within shorter time-scales of only around several nanoseconds~\cite{erhardutt2024structure}.
In contrast, in the case of $\alpha$-quartz we see sufficient time to allow an arrangement of the oxygen atoms, however, not an arrangement of the silicon atoms to stishovite.
Instead, we can only see the increase of phases like NaTiF$_4$,SnO$_2$,$P2_1/c$-structured silica and seifertite, which indicates some ordering of silicon.
However, for the occurrence of stishovite probably significantly longer times under the shock state would be required.

\subsection*{Formation of rosiaite-structured silica}

The preceding discussion highlights how kinetics significantly influence phase transformations and shock effects in silica.
The duration of the shock load and the shock temperature are critical factors affecting the outcomes observed.
Current experimental techniques and computer simulations are unable to replicate the timescales of natural impacts, as noted by DeCarli \textit{et al.} \cite{decarli_2002}. Consequently, comparing these results with observations from natural impact sites remains challenging.
However, a novel approach to mimic the effects of a prolonged load has been to utilize rapid cold compression experiments under non-hydrostatic conditions \cite{otzenEvidenceRosiaitestructuredHighpressure2023, kingma1996synchrotron}. In these experiments  $\alpha$-quartz predominantly transforms into rosiaite-type structured silica, rather than the d-NiAs-type structure observed in recent shock experiments (Tracy et al.) and in our simulations. 

While computer simulations on the timescale of seconds remain out of reach, supporting these experiments is feasible by analyzing the stress and strain conditions necessary for the formation of phases that are typically unstable under hydrostatic conditions. Recent computer simulations have proposed a direct, diffusionless transition of quartz to rosiaite-structured silica \cite{tsuchiyaNewHighpressureStructure2022}. Both, the experimental and simulation data, indicate that while the principal compression direction is parallel to the $c$-axis of $\alpha$-quartz, additional compression perpendicular to the $c$-axis is required for this transition \cite{tsuchiyaNewHighpressureStructure2022}. To comprehensively understand the strain conditions leading to this transition, we systematically strained the unit cell of $\alpha$-quartz parallel and perpendicular to the $c$-direction to various ratios and magnitudes. At each strain state, the structure was relaxed, subjected to thermal treatment and relaxed again (see Fig. 4a). Finally, the resulting structure was identified using the DG-CNN classifier. 

Depending on the ratio of strains parallel and perpendicular to the $c$-axis, the structure of $\alpha$-quartz is retained for strains of up to 25~\% and 16~\% parallel and perpendicular to the $c$-axis, respectively (Fig. \ref{fig:phases_strain}b). A transition to rosiaite-structured silica is observed for strains of 25 to 40~\% along the $c$-axis and 5 to 20 \% perpendicular to the $c$-axis. At strain conditions outside of this stability field, on the contrary, amorphous silica can be observed predominantly (See also Supplementary Fig. S5). Depending on the strain conditions, the DG-CNN identifier indicates local similarities of the resulting structures with other metastable high-pressure polymorphs, i.e. d-NiAs, SnO$_2$ and CaCl$_2$ type silica, and silica crystallizing in the space groups $C2$ and $P3_2$. Details on the identified high-pressure polymorphs can be found in Supplementary Fig. S3--S5. 

To analyse the probability for the occurrence of a transition from quartz to rosiaite-structured silica, we performed solid-state nudge elastic band (SS-NEB) calculations. These calculations enable to analyse the transition path way between two different structures with different lattice parameters. We fixed strains perpendicular to the $c$-axis and scaled $\alpha$-quartz and rosiaite-structured silica to the same lattice parameters in these directions. In $c$-direction we applied an uniaxial stress. Under these external constraints we investigate possible transition paths also considering atoms in $\alpha$-quartz may move to different positions in rosiaite-structured silica. Each path way corresponds to a \textit{blurred} line in Fig. \ref{fig:phases_strain}c-d, while the \textit{bold} line corresponds to the minimum energy path. Using strains of 10~\% perpendicular to the $c$-axis, we observe that the rosiaite-structured silica is more stable then $\alpha$-quartz when the compression exceeds approximately 15~GPa (Fig. \ref{fig:phases_strain}c). At these conditions, the phases are separated by a relatively low energy barrier of 150~meV/atom (bold line Fig. \ref{fig:phases_strain}c), which further decreases at increasing pressures. The results indicate that the transition from quartz to rosiaite-structured silica can occur very easily at room temperature when pressures of approximately 15~GPa are exceeded, which is in good agreement with experiments~\cite{otzenEvidenceRosiaitestructuredHighpressure2023}. 

On the contrary, if quartz is compressed along the $c$-axis without compressive strains in the perpendicular direction, rosiaite-structured silica becomes stable at pressures of approximately 40~GPa (Fig. \ref{fig:phases_strain}d). Additionally, the calculations yield intermediate minima on all transition paths. The structure attained at one of these minima indicates a high degree of disorder. In agreement with Fig. \ref{fig:phases_strain}b, this means that quartz amorphizes if compressed along the $c$-axis without additional compression in the perpendicular direction, whereas rosiaite-structured silica remains absent. Subsequent crystallization of the amorphous solid to rosiaite-structured silica is unlikely due to the high energy of rosiaite-structured silica compared to the energy of other high-pressure polymorphs (Fig. \ref{fig:overview}a). 

The strain conditions promoting amorphization of quartz might also be responsible for the amorphization seen in the shock simulations, where lateral strains were absent during shock compression along the $c$-axis. 
During natural impacts, on the contrary, strain conditions are strongly heterogeneous due to the random crystallographic orientations of quartz grains in rocks and locally varying interactions to neighboring crystals.

\section*{Conclusion}

In this work, we performed shock simulations of $\alpha$-quartz using a novel machine-learning interatomic potential. 
In the first part of this work, we show that $\alpha$-quartz nearly completely amorphizes under shock before it crystallizes within the d-NiAs structure. 
This observed d-NiAs structure is found to deviate from the perfect d-NiAs structure in a way that the silicon atoms are not completely randomly distributed on their sublattice. 
Instead, there is a certain ordering of the silicon atoms within the octahedral voids, which is also reflected in the appearance of small regions with other structures like seifertite, NaTiF$_4$-,~SnO$_2$-,~and $P2_1/c$-structured silica. 
In the second part, we analysed the transformation behaviour of $\alpha$-quartz into rosiaite-structured silica. 
Indeed, we found that for the diffusionless transformation not only compression along the $c$-axis direction is required, but additional compressive strain perpendicular to the main compression direction. 
This perpendicular compressive strain occurs in static compression experiments, where rosiaite-structure silica has been observed. 
In contrast, in cases where no additional lateral compression is applied, we show that it will be always more favorable for $\alpha$-quartz to amorphize. 
This work shows the high relevance of time scales as well as different strain conditions for the transformation behavior of silica. 
While conditions might seem similar, different time scales or strain states can result in a big difference in the observed phases. 
This variability might also explain the wide range of metastable silica polymorphs and the diverse shock effects documented in experiments and in natural samples from impact craters.
We demonstrate that complementing experimental results with machine-learning driven atomistic simulations offers a valuable strategy for gaining a more detailed understanding of the atomic structural responses.

\section*{Methods}

For all calculations despite the DFT calculations we used the machine-learning interatomic potential from Ref. \citenum{erhard2024modelling}. For data visualisation and processing we used \texttt{ovito} \cite{stukowskiVisualizationAnalysisAtomistic2010}, the \texttt{ase} \cite{larsenAtomicSimulationEnvironment2017} package and \texttt{pymatgen} \cite{ongPythonMaterialsGenomics2013}. Molecular dynamics and molecular statics simulations have been performed using \texttt{LAMMPS} \cite{thompsonLAMMPSFlexibleSimulation2022a}. 

\subsection*{Energy volume curves}
All energy volumes curves are calculated under hydrostatic conditions allowing to change the box shape and the atomic positions.
DFT references of the energy-volume curves of SnO$_2$-,NaTiF$_4$ and $P2_1/c$-structured silica have been determined with VASP \cite{kresseEfficiencyAbinitioTotal1996,kresseEfficientIterativeSchemes1996} using the project augmented-wave method \cite{blochlProjectorAugmentedwaveMethod1994,kresseUltrasoftPseudopotentialsProjector1999} with the SCAN exchange-correlation functional \cite{sunStronglyConstrainedAppropriately2015} and an energy-cutoff of 900 eV and a k-spacing of 0.23~\si{\angstrom}$^{-1}$. These are the same settings, which have been used for the machine-learning potential training data \cite{erhardMachinelearnedInteratomicPotential2022,erhard2024modelling}. 
Additionally, we calculated the energy-volume curves with the machine-learning potential for the mentioned structures and also for the d-NiAs structure. 
Since d-NiAs-structured silica is not uniquely defined, we took one structural sample from Ref. \citenum{erhardutt2024structure} and created additionally 9 further samples. All samples have a size of 3000~atoms. They have been created by distributing silicon randomly over the octahedral voids, however, under the constraint that no oxygen atom has less than one or more than five bonds.

\subsection*{Shock simulations}

Shock simulations have been performed using \texttt{LAMMPS}~\cite{thompsonLAMMPSFlexibleSimulation2022a} and the Hugoniostat method (\texttt{fix nphug})~\cite{ravelo2004constant}. We used a temperature damping value of 20~ps and a pressure damping value of 20~ps as input for the \texttt{fix nphug}. Initially, we equilibrated the system using NPT ensemble \cite{shinoda2004rapid} for 10~ps at 300~K. Afterwards we performed the shock simulation for a time of 60~ns. The simulation cell contains 518,400 atoms.

\subsection*{Dynamic graph convolutional neural network structure identification}
We used the machine-learning base structure identification algorithm for silica from Ref. \cite{erhardutt2024structure} based on 64 next neighbors as input. This model is available from \texttt{github} \cite{uttNnn911MLSI2022}. This machine-learning model is using a DG-CNN as classifier and is trained to differentiate between 25 different silica phases, which include also the amorphous phase, the melt, and additionally many hypothetical high-pressure structures. A list of all structures, which can be identified by this method is given for our deformation analysis of quartz (see Supplementary Figure S3-S5). The classifier gives a score for each structure type it was trained to. Correspondingly the structure of an atoms is identified to be the one with the highest score. However, it cannot identify structures, which it was not trained to. Instead, it will wrongly classify this unknown structures with one of the 25 other known structures. 

\subsection*{X-ray diffraction calculations}
X-ray diffraction intensity of the shock simulation have been calculated using the Debye formula~\cite{debyeZerstreuungRontgenstrahlen1915},

\begin{equation}
    I(q)= \sum_{i,j}f_i(q)\cdot f_j(q) \cdot \dfrac{sin(q\cdot r_{ij})}{q\cdot r_{ij}},
\end{equation}

where $q$ is the scattering vector, $f_a(q)$ is the atomic scattering factor of atom $i$ and $r_{ij}$ is the distance between atom $i$ and $j$. The scattering vector $q$ is given by,

\begin{equation}
    q = 4\pi sin(\Theta /\lambda),
\end{equation}

with the diffraction angle $\Theta$ and the wavelength $\lambda$. For our calculations we used the radial distribution function calculated from \texttt{ovito} and neglected periodic boundary conditions.
For the small-scale pristine crystals we used the functionalities implemented in \texttt{pymatgen} to evaluate the x-ray diffraction intensity. 
In all case we assume that the wavelength distribution is equivalent to the wavelength distribution in the experimental work by Tracy \textit{et al.} \cite{tracyStructuralResponseAquartz2020}.

\subsection*{Deformation calculations}
We used as input $\alpha$-quartz cells under no pressure with supercell size between $1\times 1\times 1$ and $8\times 8\times 8$. 
These cells have been deformed by using the following strain state,
\begin{equation}
    \epsilon= \begin{pmatrix}
                    \epsilon_{xx} & 0 & 0 \\
                    0 & \epsilon_{xx} & 0 \\
                    0 & 0 & \epsilon_{zz} \\
    \end{pmatrix}.
\end{equation}
After deformation we applied the following protocol, to allow a transformation of the structure:
\begin{itemize}
    \item Relaxation of the atoms
    \item Heating from 10~K to 1000~K within 10~ps
    \item Annealing at 1000~K for 10~ps
    \item Cooling from 1000~K to 10~K within 10~ps
    \item Relaxation of the atoms
\end{itemize}

All steps have been performed under a constant box size, to keep the strains fixed. 
Structure identification have been performed using \texttt{spglib} \cite{togoTextttSpglibSoftware2018} and the DG-CNN structure identifiaction (see above). 
Results for the \texttt{spglib} analysis are shown in Supplementary Fig. S3a.
The results of the DG-CNN for rosiaite-structured silica are shown in Supplementary Fig. S3b (for all supercell sizes separately) and for all other structures averaged over all supercell sizes in Supplementary Fig. S4-S5.

\subsection*{Solid-state nudge-elastic-band calculations}
The solid state nudge-edge elastic band method as implemented in the \texttt{FD-NEB} code~\cite{gao_groupFDNEB2022,ghasemiNudgedElasticBand2019} has been used. This code is based on the \texttt{tsase} code~\cite{tsase,sheppardGeneralizedSolidstateNudged2012,xiaoMechanismCaIrOPostperovskite2013}.
We used Cauchy stresses for the calculation and set the \textit{weight} parameter to a value of 100, to decrease the box fluctuation during the NEB. 
As input cell, we used $\alpha$-quartz ($2\times 2\times 2$ supercells) strained to different states using the following strain tensor,
\begin{equation}
    \label{eq:strain}
    \epsilon= \begin{pmatrix}
                    \epsilon_{xx} & 0 & 0 \\
                    0 & \epsilon_{xx} & 0 \\
                    0 & 0 & 0\\
    \end{pmatrix}.
\end{equation}
The lattice parameters in $c$-direction is subsequently determined by relaxation of position of the atoms, while simultaneously allowing the box length in $c$-direction to change. 
During this minimization we apply the external pressure $\sigma_{zz}$.
The final states of the rosiaite phase are determined similarly. 
Since oxygen from $\alpha$-quartz can move to different position within rosiaite-structured silica, we used several rosiaite-cells, which have been generated in our deformation calculations. 
We used all $2\times 2 \times 2$ supercells, which have been generated from a strain $\epsilon_{zz}=0.27$ and have a perfect $P\overline{3}1m$ symmetry. 
These supercells have been scaled to the same lattice parameters in $a$ and $b$ direction as of the $\alpha$-quartz input cell. 
This enables to vary only the $c$ parameter during the NEB, while keeping all other lattice parameters fixed.
As for $\alpha$-quartz the $c$-parameter of the rosiaite cell is determined by structural relaxation. 

The non-hydrostatic enthalpy in our calculations is given by \cite{xiaoMechanismCaIrOPostperovskite2013},

\begin{equation}
    \Delta H = \Delta E + \Omega \sigma^{\text{text}}\cdot \epsilon^{\text{ref}},
\end{equation}

where $\Delta E$ is the energy difference between the initial state and the image, $\Omega$ is the supercell volume, $\sigma^{\text{ext}}$ is the externaly applied pressure and $\epsilon^{\text{ref}}$ is the strain relative to the initial image. Note that $\epsilon^{\text{ref}}$ is not equal to Eq. \ref{eq:strain}. Since deformation is only allowed in $c$-direction with respect to the reference image, $\epsilon^{\text{ref}}$  contains only the $\epsilon_{zz}$ component while all other components are zero.

\section*{Data availability}

Input files for MD simulations as well as selected snapshots and the results from deformation simulations will be made available on Zenodo under the doi 10.5281/zenodo.11184490 upon publication.

\normalsize

\bibliography{literature}

\section*{Acknowledgment}

L.C.E. acknowledges helpful discussion with Marcel Sadowski and Niklas Leimeroth. J.R. and K.A. acknowledge support by the Deutsche Forschungsgemeinschaft (DFG, Grant no. 405621137, 405621160).

This work was performed on the HoreKa supercomputer funded by the
Ministry of Science, Research and the Arts Baden-Württemberg and by
the Federal Ministry of Education and Research.
The authors gratefully acknowledge the computing time provided to them on the high-performance computer Lichtenberg at the NHR Centers NHR4CES at TU Darmstadt. This is funded by the Federal Ministry of Education and Research, and the state governments participating on the basis of the resolutions of the GWK for national high performance computing at universities (www.nhr-verein.de/unsere-partner).

\section*{Author contributions}
L.C.E. performed all calculations. J.R and K.A. supervised the calculations. All authors contributed substantially to the design of the research.

\section*{Competing interests}
The authors declare no competing interests.

\end{document}


\title{Supplementary Information: Understanding phase transitions of $\alpha$-quartz under dynamic compression conditions by machine-learning driven atomistic simulations}

\author{Linus C. Erhard}
\email{erhard@mm.tu-darmstadt.de}
\affiliation{Institute of Materials Science, Technische Universit\"a{}t Darmstadt, Otto-Berndt-Strasse 3, D-64287 Darmstadt, Germany}

\author{Christoph Otzen}
\affiliation{Institute of Earth and Environmental Sciences, University of Freiburg, Hermann-Herder-Strasse 5, 79104 Freiburg, Germany}

\author{Jochen Rohrer}
\affiliation{Institute of Materials Science, Technische Universit\"a{}t Darmstadt, Otto-Berndt-Strasse 3, D-64287 Darmstadt, Germany}

\author{Clemens Prescher}
\affiliation{Institute of Earth and Environmental Sciences, University of Freiburg, Hermann-Herder-Strasse 5, 79104 Freiburg, Germany}

\author{Karsten Albe}
\email{albe@mm.tu-darmstadt.de}
\affiliation{Institute of Materials Science, Technische Universit\"a{}t Darmstadt, Otto-Berndt-Strasse 3, D-64287 Darmstadt, Germany}

\maketitle

\textbf{This pdf file includes:}

Supplementary Figure 1-5

Supplementary References

\begin{figure*}
    \centering
    \includegraphics[width=13cm]{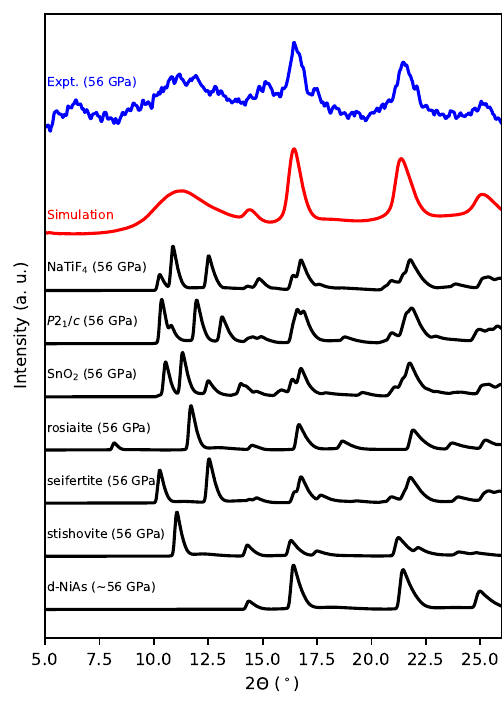}
    \caption{Theoretical x-ray pattern of various silica polymorphs, which have been compressed to 56~GPa hydrostatically compared to the simulation results from Fig. 3 and experimental results from Ref. \cite{tracyStructuralResponseAquartz2020}. For the calculations we assumed the same distribution of wavelengths as in the experimental work.}
    \label{sfig:XRD}
\end{figure*}

\begin{figure*}
    \centering
    \includegraphics[width=13cm]{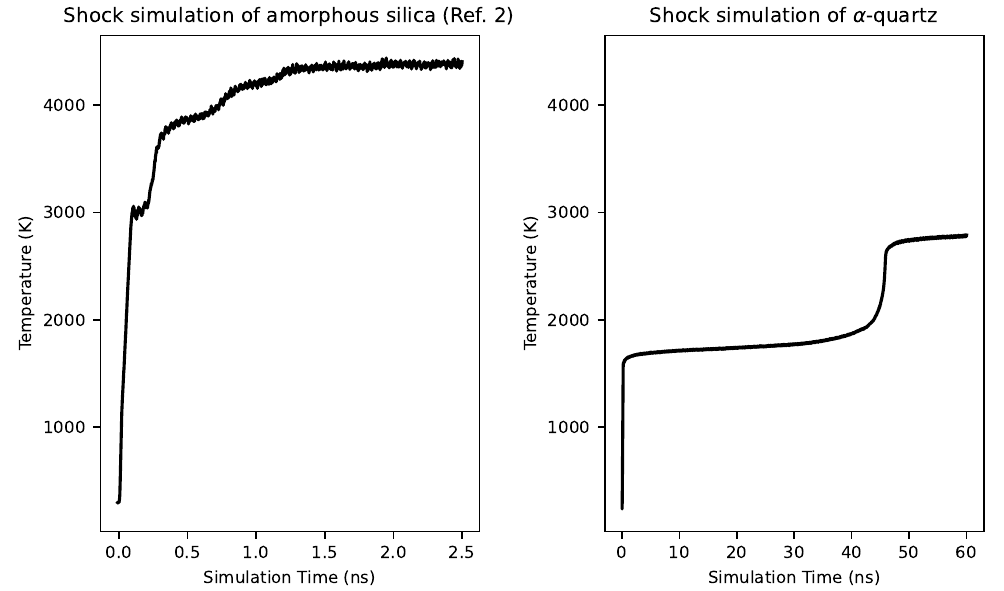}
    \caption{Temperature profiles over runtime for a shock simulation of amorphous silica (taken from Ref. \citenum{erhardutt2024structure}) under an isostatic pressure of 50~GPa compared to the temperature profile of a shock simulation of $\alpha$-quartz under uniaxial shock along the z-direction under 56~GPa (this work).}
    \label{sfig:temp}
\end{figure*}

\begin{figure*}
    \includegraphics[width=17cm]{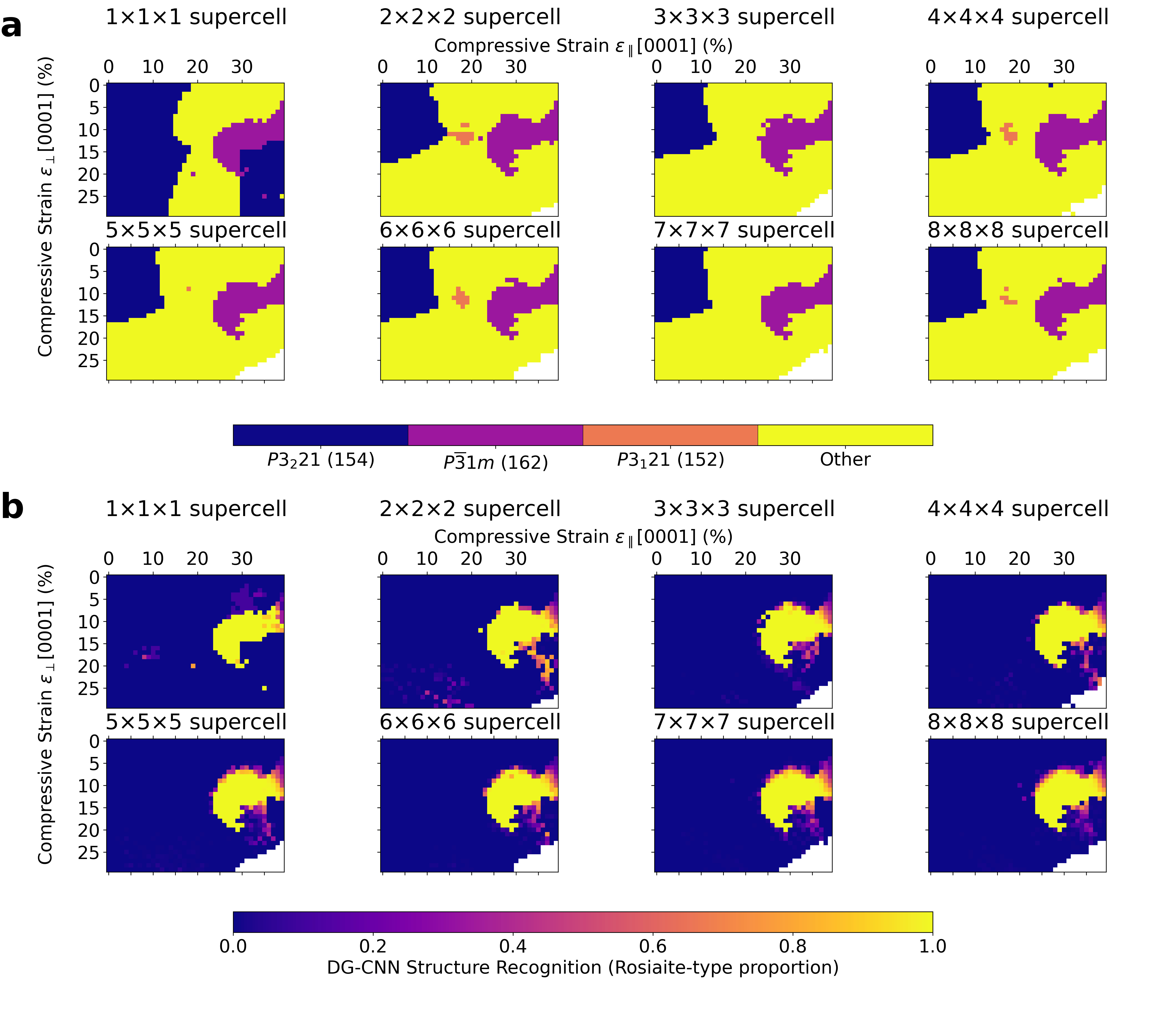}
    \caption{(a) Symmetry analysis of the structures generated by the procedure shown in Fig.~4a in the main article for different supercell sizes. Both $\alpha$-quartz and rosiaite occur independently of the size of the supercell. (b) DG-CNN structure identification analysis on the same structure only for rosiaite-type silica. The brighter a spot the higher the fraction of atoms, which have been recognized as rosiaite. White pixels correspond to a failure of the simulation under too extreme conditions.}
    \label{sfig:structID1}
\end{figure*}

\begin{figure*}
    \includegraphics[]{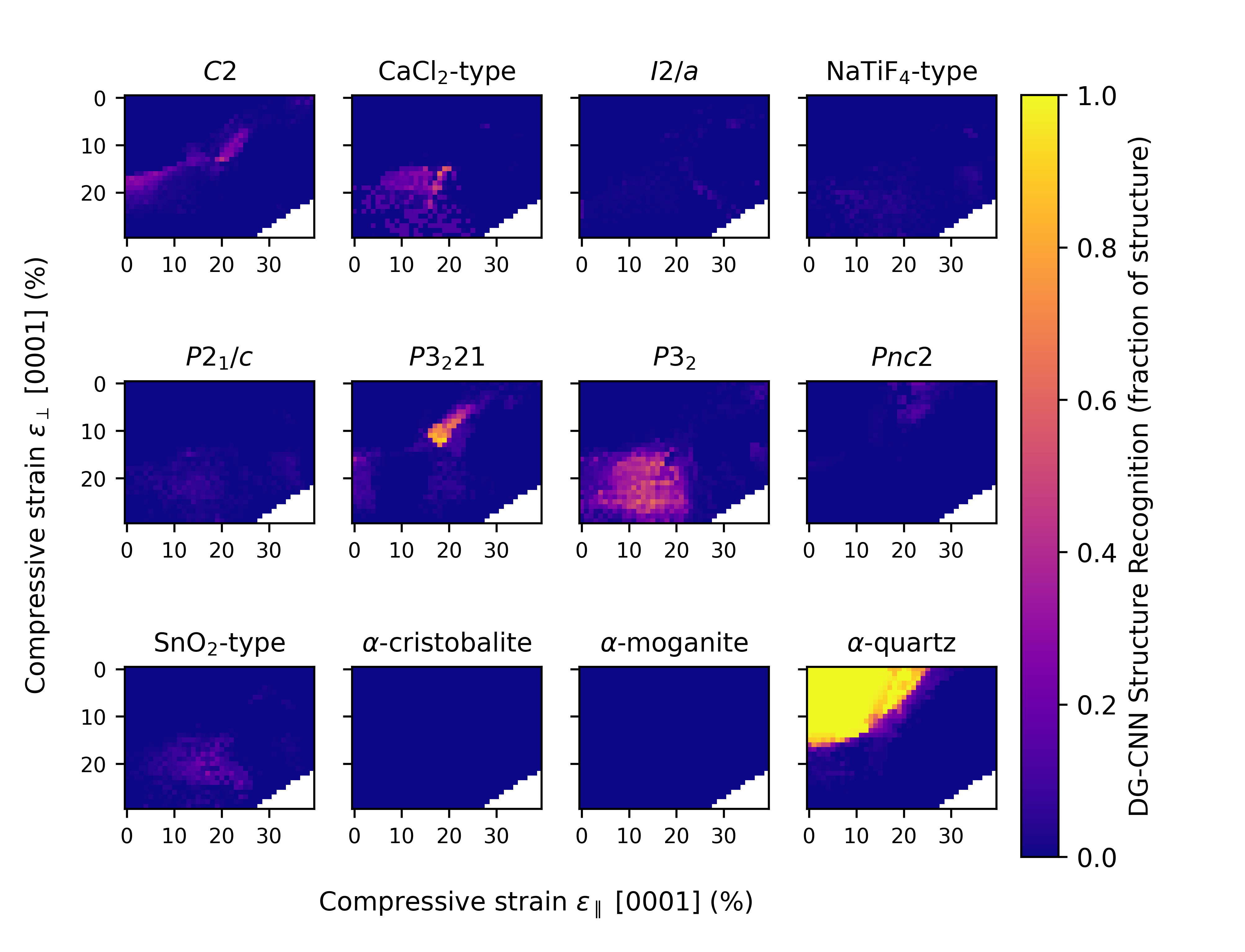}
    \caption{DG-CNN structure identification analysis of several polymorphs for the structures generated by the procedure shown in Fig.~4a. We averaged over all supercell sizes. Details about the structures $C2$~\cite{choudhuryInitioStudiesPhonon2006}, $I2/a$~\cite{tseHighpressureDensificationAmorphous1992}, $P2_1/c$~\cite{teterHighPressurePolymorphism1998}, $P3_221$~\cite{badroTheoreticalStudyFivecoordinated1997}, $P3_2$~\cite{wentzcovitchNewPhasePressure1998}, $Pnc2$~\cite{svishchevOrthorhombicQuartzlikePolymorph1997} can be found in the corresponding publications. Brighter points correspond to a higher fraction of this structure under given strain conditions. White pixels correspond to a failure of the simulation under too extreme conditions.}
    \label{sfig:structID2}
\end{figure*}

\begin{figure*}
    \includegraphics[]{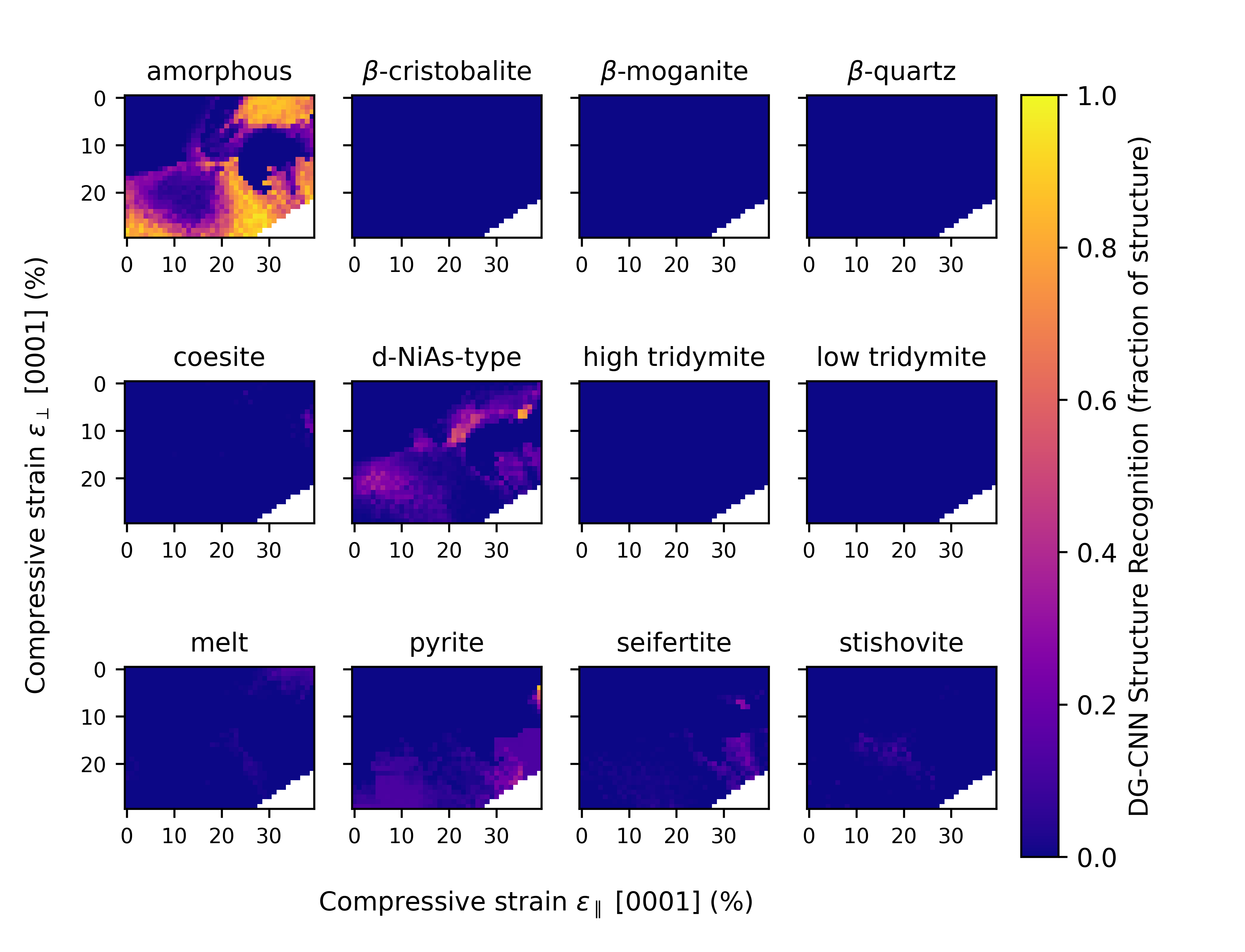}
    \caption{DG-CNN structure identification analysis of several polymorphs for the structures generated by the procedure shown in Fig.~4a. We averaged over all supercell sizes. Brighter points correspond to a higher fraction of this structure under given strain conditions. White pixels correspond to a failure of the simulation under too extreme conditions.}
    \label{sfig:strucID3}
\end{figure*}

\clearpage

\bibliography{Supplementary_literature}